\documentclass[%
 aip,
 amsmath,amssymb,
reprint,%
]{revtex4-1}
\usepackage{graphicx}
\usepackage{dcolumn}
\usepackage{bm}

\usepackage[normalem]{ulem}
\usepackage{xcolor}
\usepackage[utf8]{inputenc}
\usepackage[T1]{fontenc}
\usepackage{mathptmx}
\usepackage{url}

\begin{document}

\preprint{AIP/123-QED}


\title[]{Broadband, millimeter-wave anti-reflective structures on sapphire ablated with femto-second laser}

\author{R. Takaku}
\email{takaku@ac.jaxa.jp}
\affiliation{Department of Physics, University of Tokyo, 7-3-1, Hongo, Bunkyo-ku, Tokyo 113-0033, Japan}

\author{S. Hanany}
\affiliation{School of Physics and Astronomy, University of Minnesota/Twin Cities, Minneapolis, MN, 55455, USA}

\author{H. Imada}
\affiliation{Kavli Institute for the Physics and Mathematics of the Universe, 5-1-5 Kashiwa-no-Ha, Kashiwa City, Chiba 277-8583, Japan}

\author{H. Ishino}
\affiliation{Okayama University, 3-1-1, Tsushimanaka, Kita-ku, Okayama City, Okayama, 700-8530, Japan}

\author{N. Katayama}
\affiliation{Kavli Institute for the Physics and Mathematics of the Universe, 5-1-5 Kashiwa-no-Ha, Kashiwa City, Chiba 277-8583, Japan}

\author{K. Komatsu}
\affiliation{Okayama University, 3-1-1, Tsushimanaka, Kita-ku, Okayama City, Okayama, 700-8530, Japan}

\author{K. Konishi}
\affiliation{Institute for Photon Science and Technology, the University of Tokyo, 7-3-1, Hongo, Bunkyo-ku, Tokyo 113-0033, Tokyo, Japan}

\author{M. Kuwata-Gonokami}
\affiliation{Department of Physics, University of Tokyo, 7-3-1, Hongo, Bunkyo-ku, Tokyo 113-0033, Japan}

\author{T. Matsumura}
\affiliation{Kavli Institute for the Physics and Mathematics of the Universe, 5-1-5 Kashiwa-no-Ha, Kashiwa City, Chiba 277-8583, Japan}

\author{K. Mitsuda}
\affiliation{Institute of Space and Astronautical Science, Japanese Aerospace Exploration Agency, 3-1-1, Yoshinodai, Chuo-ku, Sagamihara City, Kanagawa, 252-5210, Japan}

\author{H. Sakurai}
\affiliation{Institute for Photon Science and Technology, the University of Tokyo, 7-3-1, Hongo, Bunkyo-ku, Tokyo 113-0033, Tokyo, Japan}

\author{Y. Sakurai}
\affiliation{Kavli Institute for the Physics and Mathematics of the Universe, 5-1-5 Kashiwa-no-Ha, Kashiwa City, Chiba 277-8583, Japan}

\author{Q. Wen}
\affiliation{School of Physics and Astronomy, University of Minnesota/Twin Cities, Minneapolis, MN, 55455, USA}

\author{N. Y. Yamasaki}
\affiliation{Institute of Space and Astronautical Science, Japanese Aerospace Exploration Agency, 3-1-1, Yoshinodai, Chuo-ku, Sagamihara City, Kanagawa, 252-5210, Japan}

\author{K. Young}
\affiliation{School of Physics and Astronomy, University of Minnesota/Twin Cities, Minneapolis, MN, 55455, USA}

\author{J. Yumoto}
\affiliation{Department of Physics, University of Tokyo, 7-3-1, Hongo, Bunkyo-ku, Tokyo 113-0033, Japan}
\affiliation{Institute for Photon Science and Technology, the University of Tokyo, 7-3-1, Hongo, Bunkyo-ku, Tokyo 113-0033, Japan}

\date{\today}

\begin{abstract}
We designed, fabricated, and measured anti-reflection coating (ARC) on sapphire that has 116\% fractional bandwidth and transmission of at least 97\% in the millimeter wave band. The ARC was based on patterning pyramid-like sub-wavelength structures (SWS) using ablation with a 15~W femto-second laser operating at 1030~nm. One side of each of two discs was fabricated with SWS that had a pitch of 0.54~mm and height of 2~mm. The average ablation volume removal rate was 1.6~mm$^{3}$/min. Measurements of the two-disc sandwich show transmission higher than 97\% between 43 and 161~GHz. We characterize instrumental polarization (IP) arising from differential transmission due to asymmetric SWS. We find that with proper alignment of the two disc sandwich RMS IP across the band is predicted to be 0.07\% at normal incidence, and less than 0.6\% at incidence angles up to 20 degrees. These results indicate that laser ablation of SWS on sapphire and on other hard materials such as alumina is an effective way to fabricate broad-band ARC. 

\end{abstract}

\maketitle

%

\section{Introduction}
\label{sec:introduction}

Sapphire, alumina, and silicon have material properties that make them appealing for use as optical elements in the millimeter and sub-millimeter (MSM) waveband, loosely defined as 30 - 3000 GHz. Compared to plastic-based materials, they have indices of refraction near 3, giving more aberration correction power per unit lens thickness~\cite{Lamb}. They have amongst the lowest absorption loss at room temperature and when cooled to cryogenic temperatures \cite{parshin1995}, and they have thermal conductance higher by factors of hundreds, making them useful for cryogenic applications~\cite{child_1973}. Sapphire has 10\% birefringence making it an ideal half-wave plate (HWP) material for polarimetric applications. 
A number of astrophysical instruments operating in the MSM waveband are using these materials~\cite{Thornton_2016,EBEX2018,SPT-3G,Kaneko2020}. 

The high index of refraction leads to high reflection loss; the average reflectance across 30\% fractional bandwidth of a 1~cm thick slab of non-birefringent sapphire at 150~GHz is 40\%. Reduction of reflection loss is achieved by applying an anti-reflection coating (ARC).  There are two generic approaches for implementing ARC:
(i) applying layers of materials with appropriately chosen intermediate indices; and (ii) machining sub-wavelength structures (SWS) on the native optical element material~\cite{Raut2011AntireflectiveCA}. 

An advantage of the SWS approach is that it does not require new materials and glues. This advantage matches well the needs of cryogenic instruments in which differences in coefficients of thermal expansion make the application of multi-layers challenging. Another advantage is that SWS give flexibility in synthesizing any index profile between free space and the substrate material.
Prescriptions exist for index profiles that give fractional bandwidths exceeding 100\% with maximal, in some sense optimal, in-band transmission~\cite{Klopfenstein}. 

Two classes of MSM instruments that require broad-bandwidths and cryogenic optical elements are those mapping the spatial polarization of the cosmic microwave background radiation (CMB)\cite{Johnson_2007,EBEX2018,SPT-3G,Simon2018,Kaneko2020,SO-OverviewGalitzki10.1117/12.2312985,cmb-s4techbookfirsted}, and others measuring the properties of Galactic dust \cite{Dober_blastpol10.1117/12.2054419,blasttngLourie10.1117/12.2314396,nika2pisano2020}. A number of these instruments have used, or plan to use, a sapphire-based HWP polarization modulator. 
To reduce detector noise the HWP is maintained at cryogenic temperatures, typically near 4~K, and to increase instrument throughput a single HWP operates over a broad range of frequencies, making a single quarter-wave layer of ARC inadequate. Examples of relevant past instruments include BLASTpol and EBEX, which had sapphire HWPs with operating bandwidth of 69\% and 109\%, respectively\cite{Dober_blastpol10.1117/12.2054419,EBEX2018}. The bandwidth of the EBEX HWP is the largest reported to date. Ongoing and future experiments include POLARBEAR2, Simons Observatory and LiteBIRD\cite{Kaneko2020,SO-OverviewGalitzki10.1117/12.2312985,sugai}. 

LiteBIRD is a Japanese-led space mission scheduled to be launched late in the next decade~\cite{sugai,YSakurai}. LiteBIRD's low frequency telescope (LFT), designed to operate between 34 and 161~GHz (a fractional bandwidth of 130\%), will have an aperture diameter of 40~cm with a HWP operating at 20~K. Currently no ARC technology on sapphire is available over this broad bandwidth. Implementing SWS ARC for the LiteBIRD HWP has motivated the developments we report in this paper.

Although SWS have advantages as ARC, their implementation on hard materials such as sapphire and alumina present fabrication challenges~\cite{Araujo2016,JAIN20021269,TUERSLEY1994377,machining_difficult}. Broad bandwidths require SWS with aspect ratios $a$ -- defined as height/pitch -- reaching four. 
Diameters of anticipated sapphire and alumina optical elements, which are reaching 80~cm\cite{AliCPTHong10.1093/nsr/nwy019}, require commercially viable machining speed.  
To overcome these challenges we demonstrated a technique to fabricate mm-wave SWS on sapphire, alumina, and silicon using laser micro-fabrication~\cite{tomo,karl,victor,tomo_ISSTT}. 
With laser ablation, which has already been used in the past to ablate these and other materials~\cite{Bonse2002,Ihlemann1995,WANG2004221}, there is no wear and tear of the machining tool. 
Even for materials that can be machined using conventional tools, laser ablation gives finer control of structure shapes, because laser spot diameters can be focused to smaller sizes than other tools. A spot diameter of 16~$\mu$m has been demonstrated at the IR~\cite{HSakurai}; smaller spot diameters are achievable in the UV.
Finer control of SWS shapes would give higher transmission over broader bandwidth.

Sch\"{u}tz et al.\ and Matsumura et al.\ (Refs.~\onlinecite{victor,tomo}) demonstrated laser-ablated aspect ratios $a =$ 
2.2 and 2.5
and heights between 700 and 800~$\mu$m, with sapphire and alumina, respectively. 
Young et al.\ (Ref.~\onlinecite{karl}) extended the technique to silicon, showing $a=4$ with height near 700~$\mu$m. To demonstrate applicability to lower frequencies and broader bandwidths, Matsumura et al.\ (Ref.~\onlinecite{tomo_ISSTT}) fabricated structures with $a=5.3$ and total height of 2.1~mm on sapphire. In all of these cases, the volume ablation rate was small, making fabrication of large samples prohibitively expensive; for example, at the rate Matsumura et al.\ (Ref.~\onlinecite{tomo_ISSTT}) quoted for the 2.1~mm tall structures on sapphire it would have taken two months to fabricate a 10~cm diameter sample. 
Further steps in the development of laser micro-fabrication of mm-wave SWS include demonstrating ablation rates that would make commercial fabrication viable with SWS aspect ratios suitable for broad bandwidth applications~\cite{wen_inprep}. In this paper we present progress in the design, fabrication, and characterization of laser-ablated SWS ARC, aiming for 130\% fractional bandwidth ARC centered on 97~GHz with improved fabrication speed. 
In Section~\ref{sec:design}, we present the design of the SWS. Sections~\ref{sec:sample} and~\ref{sec:transmission} give details of the fabrication of two sapphire samples and their transmission properties. We discuss the results and give conclusions in Sections~\ref{sec:discussion} and~\ref{sec:conclusion}.

\section{Design}
\label{sec:design}

LiteBIRD's LFT, one of three telescopes aboard the spacecraft~\cite{sugai}, will operate over a frequency range between 34 and 161~GHz. The focal plane will have about 1200 bolometric detectors tuned to nine broad-band, $\sim$25\% fractional bandwidth frequency bands.  A single HWP placed at the entrance aperture of the LFT will modulate the polarization of sky signals across the entire bandwidth. Thus, the goal is to demonstrate SWS ARC with 130\% fractional bandwidth centered on 97~GHz.

Klopfenstein (Ref.~\onlinecite{Klopfenstein}) derived a prescription for impedance matching between free space and a sample of index $n$. The prescription gives an optimal index of refraction profile and depends on a parameter $\Gamma_{m}$ that controls the trade-off between the lowest frequency of the passband and transmission ripple in-band~\cite{Grann}. All values of $\Gamma_{m}$ between 0.01 and 0.1 give average band transmission between 98\% and 99\%, and we chose $\Gamma_{m} =0.055$ as a fiducial value. With this value average transmission is near 99\%. There are several approaches to transforming the index profile to a physical shape but there is no single closed-form solution~\cite{Grann, Chen:95}. 
We employ an empirical design approach: we construct physical shapes and use effective medium theory (EMT)~\cite{EMT} and rigorous coupled-wave analysis (RCWA)~\cite{1457488,Moharam} to calculate the resulting index profile. We choose the physical shape that most closely reproduces the desired Klopfenstein index profile. RCWA calculations were carried out using DiffractMOD~\cite{diffractmod}.

\begin{figure}[t]
    \centering
    \includegraphics[width=0.5\textwidth]{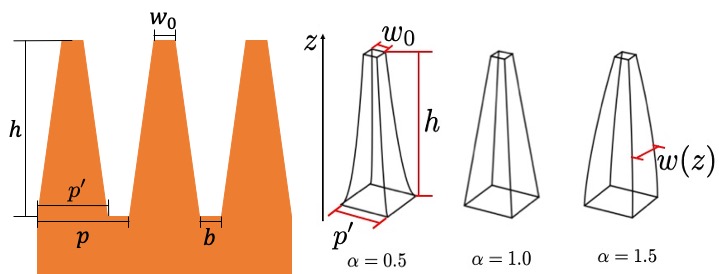}
    \caption{Geometrical parameters of pyramidal-shaped SWS. The shape of the pyramids (right panel) is controlled by a parameter $\alpha$ defined by Equation~\ref{eq:alpha}. Other parameters are the tip width $w_0$, the period $p$, the width at the bottom of the pyramid $p^\prime$, the groove width $b$, and the height $h$.
    \label{fig:Shape_parameters} }
\end{figure}

We model the SWS as a grid of identical pyramids, see Figure~\ref{fig:Shape_parameters}.  We search for a pyramid shape that closely matches the Klopfenstein index profile by constructing a shape function that gives the width of the pyramid as a function of distance $z$ from the substrate material
\begin{equation}
    w(z) = w_0 + \left\{(p-b)-w_0\right\}\left\{1-(z/h)^{\alpha}\right\},
\label{eq:alpha}
\end{equation}
where the parameters are defined in Figure~\ref{fig:Shape_parameters}. 
A value of $\alpha=1$ gives linear slopes; smaller (larger) values give concave (convex) slopes. We considered several values for $\alpha$, but in all cases used $p = 0.54$~mm and $h=2.0$ mm, giving $a=3.7$. The pitch $p$ is set by requiring that at normal incidence the smallest value of the highest pass-band frequency be $\nu_{high} = c/p \, n_{s} = 180$~GHz. Thus $p=\lambda_{high}/n_{s}=0.54$~mm, where $n_{s} = 3.06$ is the index of c-cut sapphire. Smaller value of $p$ implies shorter $\lambda_{high}$, higher $\nu_{high}$, and thus larger bandwidth. In Section~\ref{sec:non_normal} we discuss non-normal incidence. Guidance for the minimum required $h$ is obtained by considering the lowest pass-band frequency and applying the standard criterion for maximum transmission with a single ARC layer that has a uniform index $n_{l}$. In that case $n_{l}=\sqrt{n_{s}}$, and the thickness of the layer should be $t=\lambda_{0}/4n_{l}$, where $\lambda_{0}$ is the vacuum wavelength. 
For the pyramids we found empirically that $h$ should satisfy $h \gtrsim \lambda_{0}/3n_{l} = 1.9$~mm, where $\lambda_0=8.8~$mm.
We required $h \geqq 2.0$~mm.
The choice of $w_0$ was determined iteratively by trying several values. We determined that $w_0=0.1$~mm is the largest value that would still give an effective index that matches the Klopfenstein profile. Larger values produce a mismatch of the indices near $n=1$. We assumed that the deep laser ablated grooves will give $b=0$ making $p=p'$ (see Figure~\ref{fig:Shape_parameters}).

\begin{figure}[t]
    \centering
    \includegraphics[width=0.5\textwidth]{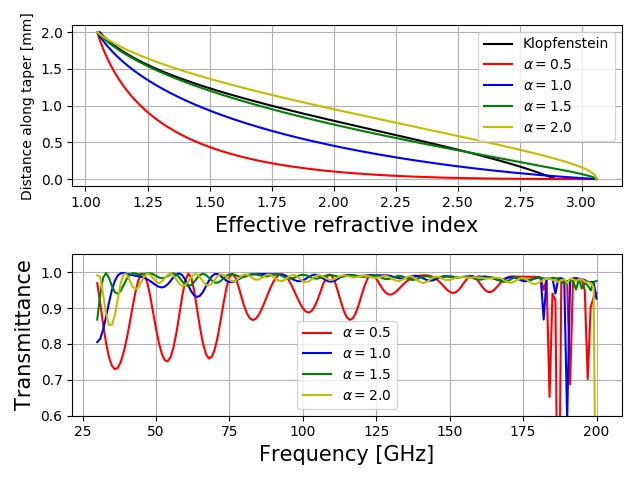}
    \caption{Top: EMT-calculated index profiles for four values of $\alpha$ (colored lines, Equation~\ref{eq:alpha}), and the Klopfenstein index profile with $\Gamma_{m} = 0.055$ (black). The other physical parameters used to calculate the index profiles are $h=2.0$ mm, $p = 0.54$~mm, $w_0=0.1$~mm, and $b=0$. Bottom: calculated transmittance as a function of frequency for SWS designs with the parameters given in the upper panel. Sharp reflection features above 180~GHz are due to the onset of diffraction. 
    \label{fig:RCWA_profile_trans} }
\end{figure}

We used second-order EMT to calculate the effective index of refraction as a function of $\alpha$, and RCWA to calculate the expected transmission as a function of frequency; see Figure~\ref{fig:RCWA_profile_trans}. A convex pyramid shape with $\alpha = 1.5$ gives an index profile that closely matches the Klopfenstein profile with the fiducial $\Gamma_{m}$. The average band transmissions for $\alpha = 1.0, \, 1.5$ and 2 are 97.7\%, 98.4\%, and 98.3\%, respectively; see Figure~\ref{fig:RCWA_profile_trans}. 
The sharp reflection features apparent in the Figure above 180~GHz are due to the onset of diffraction.

\section{Sample Preparation and characterization}
\label{sec:sample}

\subsection{Sapphire}
\label{sec:sapphire}

We procured three samples of c-cut sapphire that were cut from the same ingot, each $3.150 \pm 0.002$~mm thick and 100~mm diameter. 
We used one sample and the apparatus discussed in Section~\ref{sec:transmission} to measure the index of refraction and loss and obtained $n=3.062 \pm 0.002$ and $\tan{\delta}<1\times10^{-4}$, which is consistent with other data~\cite{Lamb}. In subsequent analysis and simulations we assume that these values are common to all three samples. 

We fabricated the SWS on a circular area of 34.5~mm diameter on one side of each of two samples, to which we refer as Sample 1 and Sample 2. We patterned only one side because LiteBIRD's LFT HWP will be a Pancharatnam multi-stack achromatic HWP~\cite{Pancharatnam1} with SWS only on the outermost surfaces. 
In Section~\ref{sec:transmission} we present transmission measurements of each sample and when they are stacked flat side on flat side. 

Using c-cut, non-birefringent sapphire, simplifies interpretation of the results. At normal incidence, which is the only experimental data we present here, any apparent birefringence is the result of asymmetry in fabrication, not of inherent asymmetry in the material, as would be the case with birefringent a-cut sapphire.

\subsection{Laser machining}
\label{sec:laser}

\begin{figure}[h]
    \centering
    \includegraphics[width = 0.4\textwidth]{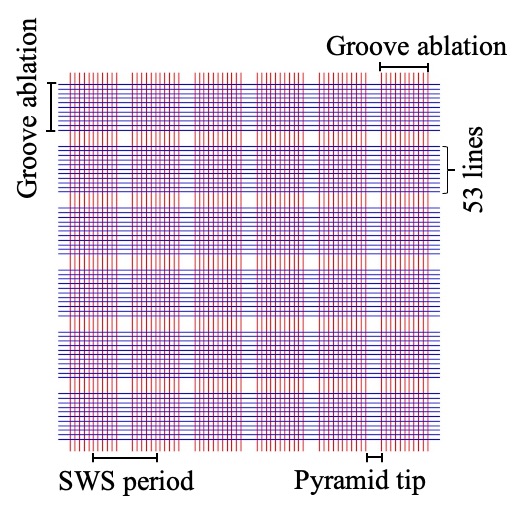}
    \caption{Sketch of the laser beam scan pattern to fabricate the SWS. Repeated ablation of groups of 53 lines produces grooves. The lines are internally spaced by less than 10 $\mu$m and the groups are separated by at least 72~$\mu$m.  Grooves fabricated in two orthogonal directions leave pyramids, which are the SWS. Layers similar to the one shown in the sketch are repeated 60 times until the desired SWS height is achieved. } 
    \label{fig:scan_strategy}
\end{figure}

The SWS are patterned using a 15~W average power femto-second laser operating at 1030~nm; the laser parameters are given in Table~\ref{tab:laser_specification}. 
The laser beam scans lines in two orthogonal directions to make grooves, and thus produce the pyramids; see Figure~\ref{fig:scan_strategy}. This scan strategy largely follows an approach we used in the past, see for example Young et al. (Ref~\onlinecite{karl}).
The focus position is set at a fixed $z=-0.75$~mm throughout the ablation. The surface of the disk at the beginning of ablation is at $z=0$ and negative values are inside the sample.

\begin{table}[h]
    \centering
    \caption{Laser Specifications}
    \begin{tabular}{c|c}\hline
        \multicolumn{2}{c}{Model:  Pharos, PH1-15W} \\ \hline
         Wavelength & 1030 nm \\
         Repetition rate &  75 kHz\\
         Pulse duration & 290 fs \\
         Pulse energy & 200 $\mu$J \\ 
         Spot diameter ($1/e^2$) & 15.5 $\mu$m \\
    \end{tabular}
    \label{tab:laser_specification} 
\end{table}

\subsection{Characterization of fabricated structures}
\label{sec:charaterization}

\begin{figure}[t]
    \centering
    \includegraphics[width=0.5\textwidth]{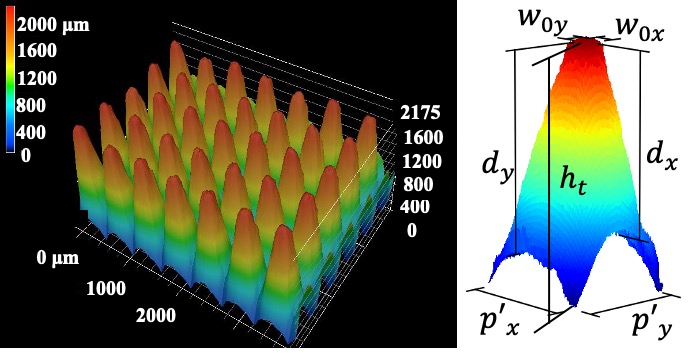}
    \caption{  Left: Confocal microscopy image of a section of the fabricated SWS. The structures are physically intact across the entire sample and the average height is 2~mm. Right: Definition of measurement parameters. Averages of 800 measurements of these parameters are given in Table~\ref{tab:geometrical_measurement_parameters_table}.   
    \label{fig:sample_shape} }
\end{figure}

The SWS were imaged using confocal imaging. A three-dimensional image of a section of the fabricated region is shown in Figure~\ref{fig:sample_shape}.   Visual inspection indicates that all SWS are physically intact with no breakage nor damaged tips. We measured 160 structures at each of five  locations in each sample, in the center and in four edge regions, and quantified several geometrical parameters as detailed in Figure~\ref{fig:sample_shape} and Table~\ref{tab:geometrical_measurement_parameters_table}. Values given in the Table are averages and standard deviations of the 800 measurements. We established a global cartesian coordinate system that was aligned with the grooves, and for each measured pyramid we defined $xz$ and $yz$ planes that intersected the center of its peak. For both planes (and for each pyramid) we quantified the widths $w_{0x}$ and $w_{0y}$ and two `saddle' heights $d_{x}$ and $d_{y}$, which quantified the depths between pyramids in the $x$ and $y$ directions. The total height $h_{t}$ is given by the depth into trenches in the diagonal direction where the laser beam ablates in both the $x$ and $y$ passes. We best-fit the slopes of the pyramids as imprinted in the $xz$ and $yz$ planes to produce measured $\alpha_{x}$ and $\alpha_{y}$.

There is good agreement between the design and fabricated values for the pitch and total height, and there is $x$-$y$ symmetry in the measured pitch. There is good repeatability between the two samples, which were fabricated several days apart. Asymmetry of 5\%-20\% is apparent at the tip of the structures, the saddle heights, and the slope values $\alpha$. This structural asymmetry could have been caused by laser beam polarization, which was not monitored; by the ablation scan strategy, for example the order in which $x$ and $y$ directions were scanned; or by laser beam projection, if the sample was not precisely normal to the beam.

It took 10.5~hours to fabricate each sample giving an average volume removal rate (AVRR) of 1.6~mm$^{3}$/min. This rate is 18 times faster than achieved in one of our earlier publications~\cite{tomo_ISSTT}. The higher rate is due to using five times higher averaged power and a more efficient sample scan pattern. Matsumura et al.\ (Ref.~\onlinecite{tomo}) report a higher AVRR of 2.2~mm$^3$/min, but with higher power (25~W) and for SWS with smaller total height of 715~$\mu$m; Schuetz et al.\ (Ref.~\onlinecite{victor}) show that AVRR increases with power and lower structure height.

\begin{figure}[h]
    \centering
    \includegraphics[width = 0.5\textwidth]{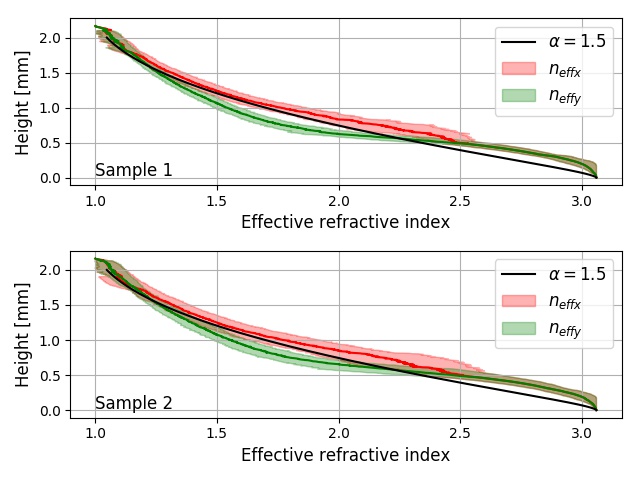}
    \caption{ The effective index of refraction calculated with EMT using the average values of the measured SWS (solid red and green), and with the design values and $\alpha = 1.5$ (solid black). The shaded regions include index profiles for all 800 measured structures.
    \label{fig:pharos8_9_EMT} }
\end{figure}

Using EMT and the average values given in Table~\ref{tab:geometrical_measurement_parameters_table} we calculated effective index profiles in $x$ and $y$ for each sample; see Figure~\ref{fig:pharos8_9_EMT}. To indicate the variance we included the index profiles for all 800 measured structures as shaded regions. As the values in Table~\ref{tab:geometrical_measurement_parameters_table} indicate, the measured profiles in $x$ match the design profile with $\alpha = 1.5$ better than the measured profiles in $y$.

\begin{table}[t]
    \centering
    \caption{
    Averages of measured geometrical parameters of 800 pyramids measured in five regions of each sample. The parameters are defined in Figure~\ref{fig:sample_shape}. The accuracy of the confocal microscope is 1~$\mu$m in the $xy$ plane and 8~$\mu$m in the $z$ direction.}
    \begin{tabular}{c|c|c|c}
         Parameter     &    Design value & Sample 1 & Sample 2  \\ 
                       &    (mm) & (mm)      & (mm) \\ \hline
        Top width $x$ ($w_{0x}$) & 0.1 & $0.14 \pm 0.01 $ & $0.14 \pm 0.01 $\\
        Top width $y$ ($w_{0y}$) & 0.1 & $0.12 \pm 0.01 $ & $0.12 \pm 0.01 $\\
        Saddle height $x$ ($d_{x}$) & & $1.51 \pm 0.04 $ & $1.52 \pm 0.04 $\\
        Saddle height $y$ ($d_{y}$) & & $1.60 \pm 0.04 $ & $1.59 \pm 0.05 $\\
        Total height ($h_{t}$) & 2.0 & $2.03 \pm 0.04 $ & $2.04 \pm 0.05 $\\
        Pitch $x$ ($p'_{x}$) & 0.54 &  $0.54 \pm 0.01 $ & $0.54 \pm 0.01 $\\
        Pitch $y$ ($p'_{y}$) & 0.54 & $0.54 \pm 0.01 $ & $0.54 \pm 0.01 $\\ \hline 
        $\alpha_x$ $^{\dagger}$ & 1.5& $1.56 \pm 0.06 $ & $1.56 \pm 0.04$\\
        $\alpha_y$ $^{\dagger}$ & 1.5 & $1.22 \pm 0.02 $ & $1.22 \pm 0.02$\\
\hline
\multicolumn{3}{l}{$^{\dagger}$ dimensionless quantity}
    \end{tabular}
    \label{tab:geometrical_measurement_parameters_table}
\end{table}

\section{Transmission}
\label{sec:transmission}

\begin{figure}[h]
    \centering
     \includegraphics[width=0.5\textwidth]{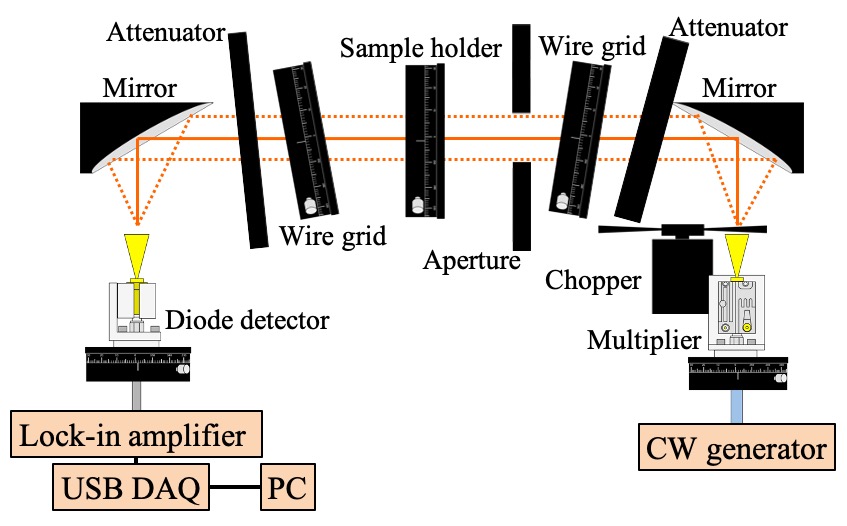}
    \caption{Schematic of the transmittance measurement setup. Additional details are given by Komatsu et al. (Ref.~\onlinecite{kkomatsu}).
    \label{fig:plane_wave_setup} }
\end{figure}

We measured the transmittance of the samples at normal incidence between 33 and 190~GHz. The apparatus, a schematic of which is shown in Figure~\ref{fig:plane_wave_setup}, consisted of a source of microwaves, an optical chopper operating at 30 Hz, two parabolic mirrors, a diode detector, and a lock-in amplifier. Between the two mirrors there was an aperture of 30 mm diameter, a sample holder, attenuators to reduce standing waves, and two always identically aligned wire grid polarizers with calculated efficiency exceeding 99\% across the bandwidth. The transmission axis of the polarizers was aligned to within 3~degrees with the $x$ axis of the samples, and the stacked samples were aligned relative to each other to within 0.5 degrees.

To further reduce the effects of standing waves our reported transmittance at each frequency is the average of two transmission measurements taken with the detector positioned at two locations spaced by $\lambda /4$ along the light path. Each transmission measurement is the ratio of power detected with the sample to the power detected without it. 

\begin{figure}
    \centering
    \includegraphics[width = 0.5\textwidth]{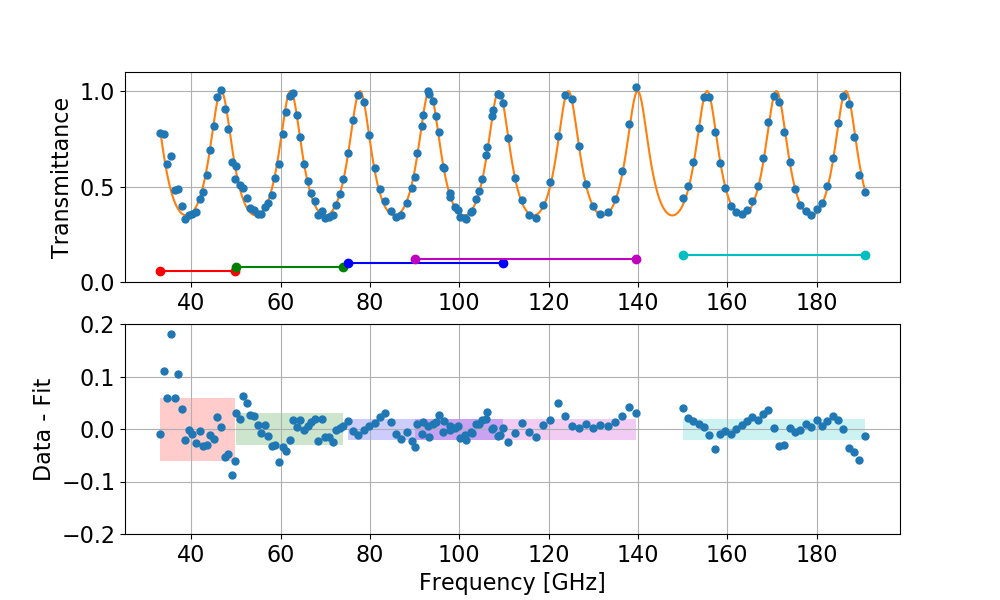}
        \caption{Top Panel: Transmittance measurement of the flat sample  (blue data) and two-parameter fit (solid orange) including the index of refraction and loss tangent. The measurement was done in five sub-bands indicated by horizontal bars. Bottom panel: residual, data minus fit, from which we infer error bars. We conservatively take the error bar per datum in each sub-band to be the RMS of the residual in the sub-band.
    \label{fig:flat_measurement} }
\end{figure}

Measurements of transmission through one of the native samples establish error estimates for subsequent data analysis. We measured transmission using different RF sources at five sub-bands between 33 and 190~GHz; see the top panel of Figure~\ref{fig:flat_measurement}. 
We fit the transmittance to a model with two free parameters, the index of refraction and loss tangent, and find the residuals; see the bottom panel of Figure~\ref{fig:flat_measurement}.  We identify systematic residuals as arising from incomplete cancellation of standing waves and 
quote the RMS of the residual as the error of transmission measurements in each sub-band. They are 6\% (33-50~GHz), 3\% (50-75~GHz), and 2\% (75-190~GHz). 

The errors quoted in Section~\ref{sec:sapphire} for the values of the index of refraction and loss were derived by fitting the transmittance data using statistical errors with magnitude per band as given above and including uncertainty on the measurement of sample thickness. The index and loss errors are dominated by the thickness measurement uncertainty.

\begin{figure*}[tbp]
    \centering
    \includegraphics[width=1.0\textwidth]{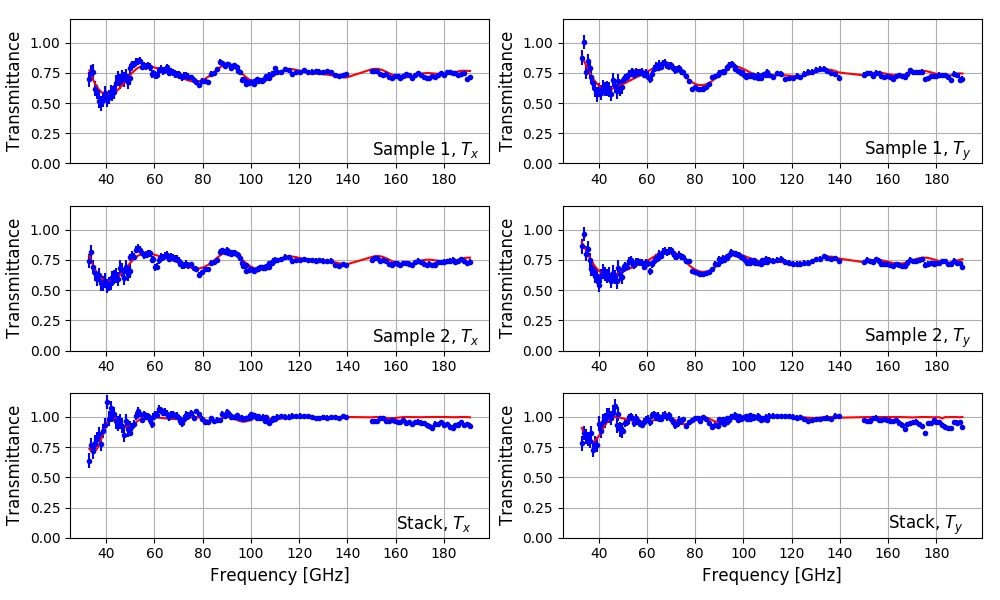}
    \caption{ Transmittance measurements $T_{x}$ and $T_{y}$ (blue data) of each sample (top four panels), of the samples stacked flat side to flat side (bottom row), and transmission predictions (red lines) based on the measured mean parameters given in Table~\ref{tab:geometrical_measurement_parameters_table}. In the bottom row, the plates are stacked with $x$ of one parallel to $x$ of the other.
    \label{fig:pharos_trans} }
\end{figure*}

We measured the transmittance of each sample and of both mechanically attached flat-side to flat-side.  The two samples were held together; no glue was used. The samples were stacked with $x$ orientations aligned parallel to each other. Each of the measurements was conducted with the transmission axis of the wire grid polarizers parallel and perpendicular to the $x$ axis of the samples. We refer to these measurements as $T_{x}$ and $T_{y}$, respectively, indicating that the incident and measured polarization states align with the $x$ and $y$ orientations of the samples. The measurements are shown in Figure~\ref{fig:pharos_trans} together with the predicted transmission, which was calculated using RCWA and was based on the average values given in Table~\ref{tab:geometrical_measurement_parameters_table} and no loss. The transmittances $T_{x}$ and $T_{y}$ of individual samples agree with predictions and are only $\sim$0.75 because they were patterned only on one side. 

For the stacked samples we find 91\% transmittance averaged between the $T_{x}$ and $T_{y}$ data for the lowest LiteBIRD frequency band between 34 and 46 GHz. For the second lowest band between 43 and 58 GHz the average is 97\%. Other frequency bands have above 98\% averaged transmittance. At frequencies above 150~GHz the transmittance of the stack is decreasing and is lower than predicted. An analysis using data over the entire bandwidth (33-190~GHz) and varying the loss tangent in the fit model gives a minimum $\chi^{2}$ for $\tan \delta < 1\times 10^{-4}$, which is consistent with measurements with the native sample. 

While determining the precise cause of lower transmittance at higher frequencies is beyond the scope of this paper, we discuss several candidates. Calculations indicate that a uniform 30~$\mu$m air gap between the stacked samples could explain the observed feature. Measurements set an upper limit of 24~$\mu$m. Ruze scattering from a flat surface with 30~$\mu$m RMS roughness could decrease transmittance to 95\% at 180~GHz~\cite{1446714}. This roughness value is close to the 30~$\mu$m uncertainty we quote for SWS height measurements (see Table~\ref{tab:geometrical_measurement_parameters_table}), which could be interpreted as an effective roughness. However, transmission measurements of the single samples indicate lower levels of effective roughness. 
Evidently, any such higher loss is not sufficiently prominent to be exhibited with significant signal-to-noise ratio with transmission measurements of the individual samples. 
A combination of these effects could be the cause of the lower transmission.

\section{Systematic Effects}
\label{sec:discussion}

Systematic asymmetry in fabrication along the two orthogonal axes, due to laser beam polarization or birefringent material properties (if such is used), leads to different effective indices of refraction along the two axes. Differential index causes differential transmission, which is a source of instrumental polarization (IP), the conversion of unpolarized to polarized light by the optical element. Even when the structures are two-dimensional symmetric there is differential reflection at non-normal incidence, as there would with any ARC scheme; differences in transmission between S and P polarization states of the incident light lead to IP. 

To quantify the IP induced by normal incidence differential transmission we use the figure of merit~\cite{radio_astronomy,karl} 
\begin{align}
    {\rm IP}(\nu) & =  \frac{ T_x(\nu) - T_y(\nu)  }{T_x(\nu) + T_y(\nu)} 
\label{eq:asymmetry_nophi} \\
    & =  \frac{ T(\phi=0^{\circ}, \nu) - T(\phi=90^{\circ}, \nu)  }{T(\phi=0^{\circ}, \nu) + T(\phi=90^{\circ}, \nu)} ,
\label{eq:asymmetry_phi1}
\end{align}
where we have introduced in the last expression the relative rotation angle $\phi$ between the $x$ orientation of the sample and the direction along which transmission is probed. IP vanishes for normal incidence light when the substrate and ARC are $z$-projected isotropic. At normal incidence  Equations~\ref{eq:asymmetry_nophi} and~\ref{eq:asymmetry_phi1} could have been written equivalently in terms of $T_{s} \equiv T(\phi=0^{o}) $ and $T_{p} \equiv T(\phi=90^{o})$, the transmissions for S and P polarization states.
 
\begin{figure}[h]
    \centering
    \includegraphics[width=0.45\textwidth]{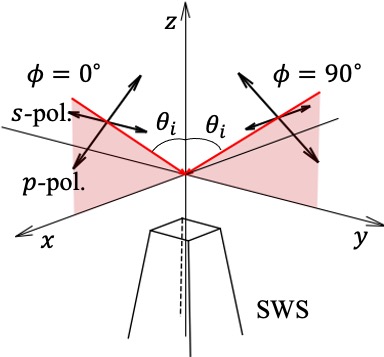}
    \caption{A diagram showing the variables of Equation~\ref{eq:asymmetry_phi2}. The transmissions $T_{p}$ and $T_{s}$ are for light polarized parallel and perpendicular to the plane of incidence, respectively.}
    \label{fig:inc_pol_iamge}
\end{figure}

At non-normal incidence and in the presence of $x$-$y$ ARC asymmetry $T_{s}$ and $T_{p}$ depend on both the angle of incidence $\theta_{i}$ and on the azimuthal angle $\phi$ of the plane of incidence, measured relative to $x$. We now have  
 \begin{equation}
    {\rm IP}(\phi,\theta_i,\nu) = \frac{ T_s(\phi,\theta_i,\nu) - T_p(\phi,\theta_i,\nu)  }{T_s(\phi,\theta_i,\nu) + T_p(\phi,\theta_i,\nu)}
\label{eq:asymmetry_phi2}
\end{equation} 
see Figure~\ref{fig:inc_pol_iamge}.

In Secs. \ref{sec:asymmetry} and \ref{sec:non_normal}, we quantify systematic effects due to asymmetry in the fabricated SWS at normal incidence (Section~\ref{sec:asymmetry}) and due to non-normal incidence (Section~\ref{sec:non_normal}), and discuss the level of mitigation provided by appropriately stacking the two-disc sandwich. We refer to `parallel' and `perpendicular' configurations, in which the two plates are stacked with their $x$ axes parallel, or with $x$ of one perpendicular to $x$ of the other, respectively. In the perpendicular configuration the quantities $T_{x}$ and $T_{y}$ refer to transmission measurements relative to the $x$ axis of the source-side sample of the stack, which was Sample 1. 

\subsection{Asymmetry of SWS at normal incidence}
\label{sec:asymmetry}

\begin{figure}[h]
    \centering
    \includegraphics[width=0.5\textwidth]{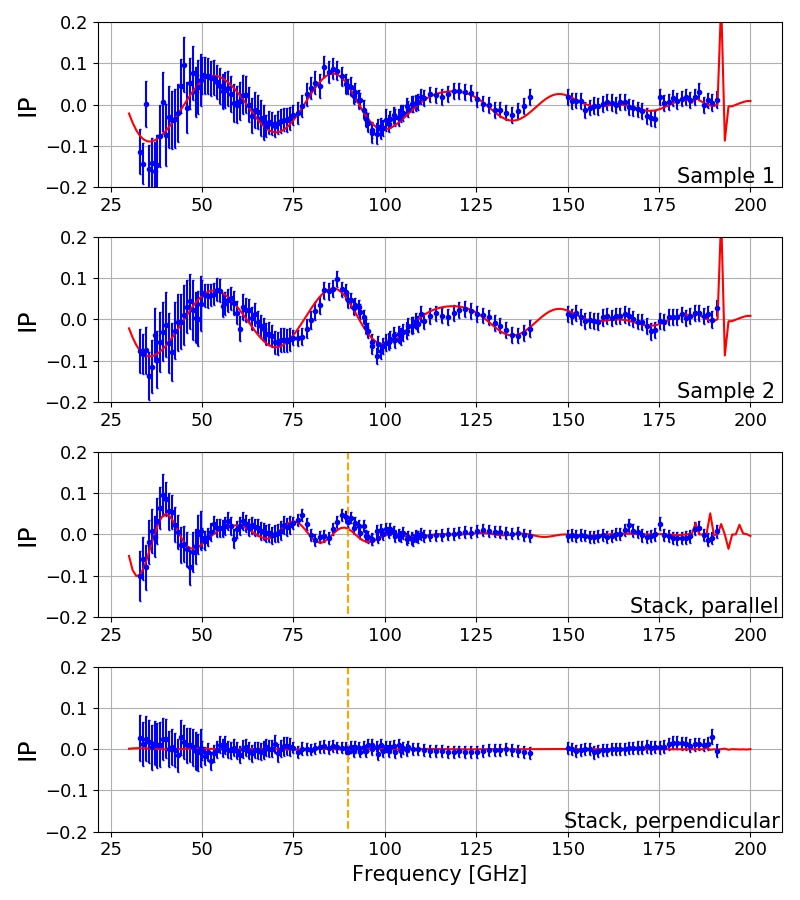}
    \caption{
    Instrumental polarization due to differential transmission IP for each of the samples and when stacked with $x$ axes parallel and perpendicular (blue data points), and RCWA predictions for IP (solid, red) based on the average measured values (Table~\ref{tab:geometrical_measurement_parameters_table}). For perpendicular stacking, calculated RMS IP for the bandwidth is 0.07\%. The vertical dashed line shows the 90~GHz frequency for which we made measurements as a function of stack rotation angle; see Figure~\ref{fig:rotation}.
    \label{fig:IP} }
\end{figure}

Figure~\ref{fig:IP} shows IP inferred from the measured data (Figure~\ref{fig:pharos_trans}) for the individual samples and when stacked in the parallel and perpendicular configurations. It also shows the calculated response (using RCWA) using the average parameters for the $x$ and $y$ orientations in Table~\ref{tab:geometrical_measurement_parameters_table}. We find values of IP reaching 10\% at frequencies below 100~GHz for the individual samples, and below 50~GHz for the parallel stacking. For the entire bandwidth the RMS IP is 2.5\%. 
At frequencies higher than 50~GHz, 
IP in the parallel stacking configuration is smaller than in the individual samples because of higher transmission and because of some averaging of the two somewhat different asymmetries. 
However, when the plates are stacked in the perpendicular configuration we find lower IP across the band as shown in the bottom panel of Figure~\ref{fig:IP}. 
The data and the RCWA prediction are consistent within measurement errors, which take into account residual standing waves (see Section~\ref{sec:transmission}).
If the errors are assumed to be statistical, we find an upper limit of 1\% on the RMS across the band. 
The RMS IP calculated based on RCWA is 0.07\%, a value that is nearly 35 times smaller than in the parallel configuration, and is below our  measurement limits.

{}

\begin{figure}[h]
    \centering
    \includegraphics[width=0.5\textwidth]{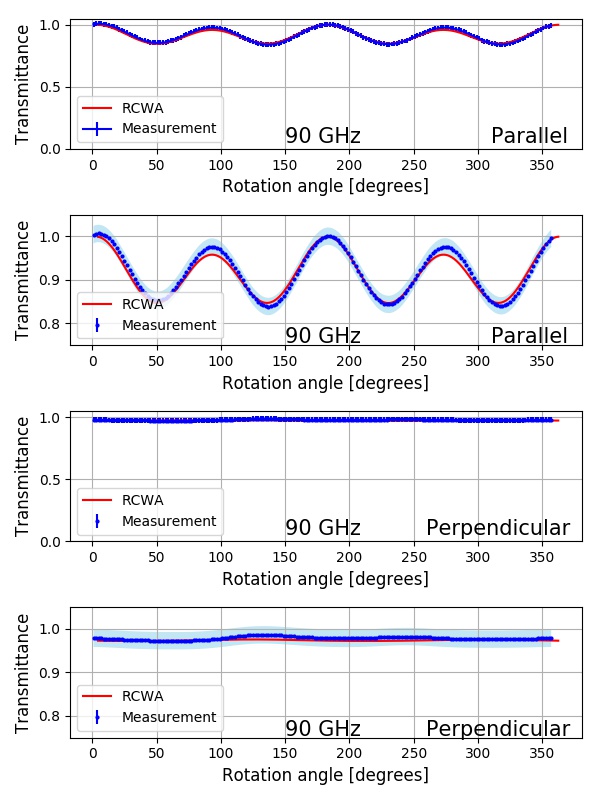}
    \caption{Stacked sample transmittance at 90~GHz (blue points) as a function of stack rotation angle $\phi$ for the parallel (top two panels) and perpendicular (bottom two panels) configurations, and RCWA predictions based on the measured parameters in Table~\ref{tab:geometrical_measurement_parameters_table} (red solid). The lower panel in each pair has a limited range for the ordinate values. Shaded regions indicate the 2\% systematic uncertainty we assign to each data point in this frequency. The statistical uncertainty per data point is 0.2\%.
    \label{fig:rotation} }
\end{figure}

Figure \ref{fig:rotation} gives data and RCWA predictions for normal incidence transmittance as a function of stacked sample rotation angle $\phi$ for the two configurations at 90~GHz. Concentrating on the parallel configuration first, the effective differential index in $x$ and $y$ is an effective stack birefringence that causes 8.6\% modulation as a function of $\phi$ with $\pi/2$ periodicity only with fully polarized incident light. 
That is, the measurements, which were conducted with fully polarized light, give
 \begin{equation}
    0.086 = \frac{ \bar{T}(\phi=0^{\circ}) - \bar{T}(\phi=45^{\circ})  }{\bar{T}(\phi=0^{\circ}) + \bar{T}(\phi=45^{\circ})}, 
\label{eq:asymmetry_phi}
\end{equation} 
with $\theta_{i} = 0^{o}$, $\nu=90$~GHz, and $\bar{T}$ indicates that we average all $T$ values at the $\phi$ indicated and with $\pi/2$ periodicity. If the experiment and simulation were conducted with partially polarized or  unpolarized light, this $\pi/2$-periodic modulation would have diminished or vanished, respectively. For the HWP, SWS using a-cut sapphire, $\pi/2$ effect also could vanish with unpolarized light. Given measurement errors in this band of frequencies as shown in Figure~\ref{fig:flat_measurement}, the $\pi/2$-periodic modulation of 0.086 has an uncertainty of 0.014. 

The IP signal due to differential transmission has $\pi$ periodicity. Given the $1\sigma = 1.4\%$ measurement uncertainty the data only give a $2\sigma$ upper limit of IP $<2.8\%$, when calculated using Equation~\ref{eq:asymmetry_phi1}. 
This value is consistent with the IP value at 90~GHz in the third panel of Figure~\ref{fig:IP}, $3 \pm 1.4\%$. This level would persist if the measurement was done with unpolarized light. 
In the perpendicular configuration both the $\pi/2$ and $\pi$-periodic modulation amplitudes, calculated using Equations~\ref{eq:asymmetry_phi} and \ref{eq:asymmetry_phi1}, are consistent with zero within measurement errors. 

The transmission data as a function of $\phi$, such as presented in Figure~\ref{fig:rotation}, reveal that measurements at only two orientations $x \, (\phi=0)$ and $y\, (\phi=90^{o})$ as shown in Figure~\ref{fig:IP} and IP calculated using Equation~\ref{eq:asymmetry_nophi} combine information from distinct physical effects. The effects correspond to, and can be quantified with higher accuracy using decomposition to Fourier harmonics\cite{Hanany:05,kkomatsu}. We fit the data of Figure~\ref{fig:rotation} to the model  
\begin{align}
    &T(\phi,\nu=90\mbox{~GHz})  =  & \nonumber \\
    &    a_0 + a_2\cos{(2\phi + 2C_2)} + a_4\cos{(4\phi + 4C_4),} &
    \label{eq:fourier}
\end{align}
and constrain the amplitudes and phases. We find 
IP$=a_2/a_0=1.6 \pm 0.2\%$ and $0.21\pm 0.22\%$, in the parallel and perpendicular configurations, respectively; and $a_4/a_0 = 7.9\pm 0.2\%$ and $0.07\pm 0.22\%$, respectively, for the effect of stack birefringence. The data in the perpendicular configuration is consistent with zero we thus quote an upper limit of 0.5\% on the values of $a_2/a_0$ and $a_4/a_0$. 
We find that (1) 
the $\pi$ and $\pi/2$ periodic modulations quantified through $a_2/a_0$ and $a_4/a_0$ are more tightly constrained compared to using Equations~\ref{eq:asymmetry_phi1} and~\ref{eq:asymmetry_phi}; (2) the modulation amplitude constraints calculated using the two techniques are consistent; (3) there is a significant detection of IP, that is, $\pi$ periodic modulation, in the parallel case but not in the perpendicular case; and (4) the relatively large 7.9\% $\pi/2$ periodic modulation measured in the parallel case is not detectable in the perpendicular case. 

With the structures as fabricated, perpendicular stacking is predicted to constrain IP to below 0.5\% at 90~GHz, a value that is limited by measurement uncertainty, not by inherent asymmetry. 

{}

\subsection{Non-normal incident light}
\label{sec:non_normal} 

Optical systems are designed to admit rays over a range of incidence planes and angles and thus  
calculations of systematic effects arising from ARC SWS asymmetry should take account of the entire range of incidence planes and angles. Including the full range amounts to averaging that gives smaller systematic effects compared to using a single incidence plane and the extreme incidence angle. In the discussion below we analyze the effects of non-normal incidence angles {\it without} any such averaging, and therefore the quantitative values should be understood as upper limits. The analysis relies solely on RCWA calculations of the perpendicular configuration; we have already established that this configuration has smaller level of systematic effects. 

\begin{figure}[h]
    \centering
    \includegraphics[width = 0.48\textwidth]{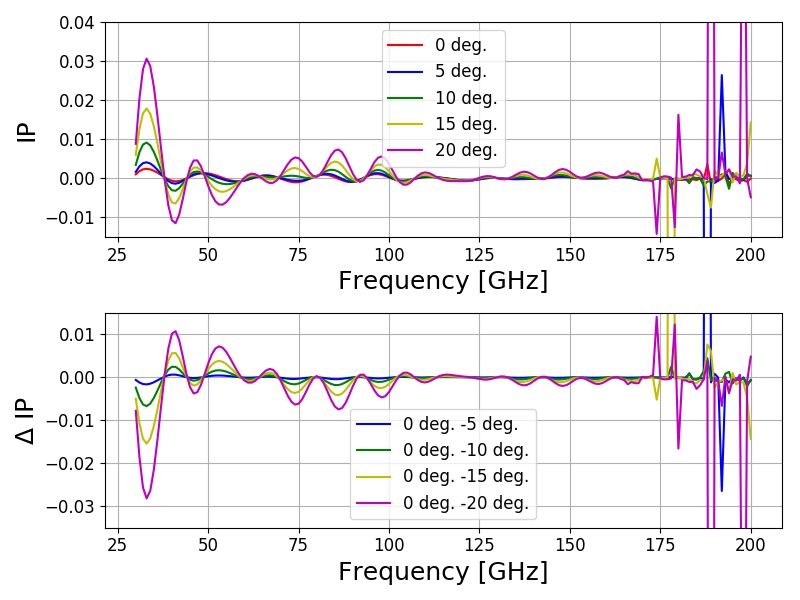}
    \caption{RCWA prediction of IP$(\phi=0)$ for various incidence angles (upper panel), and the difference of non-normal incidence spectra with normal incidence spectra (lower panel), all in the case of perpendicular stacking configuration.
    \label{fig:pharos89_inc} }
\end{figure}

The upper panel of Figure \ref{fig:pharos89_inc} shows calculated IP$(\phi=0)$ as a function of frequency for normal incidence and for the 5, 10, 15 and 20 degrees incidence angle.  For normal incidence the calculation is identical to the one shown in the lower panel in Figure~\ref{fig:IP}. The lower panel shows the difference between transmission at normal incidence and at other angles, again for $\phi=0$. For LiteBIRD's maximum planned incidence angle of 15 degrees RMS IP across the band is 0.4\%, an increase of 0.3\% relative to normal incidence. At 20~degrees incidence, RMS IP increases to 0.6\%. For $\phi=90$ degrees, RMS IP values at 15 and 20 degrees incidence are smaller by 0.1\%.  

Onset of diffraction is apparent at lower frequencies with larger incidence angles. For non-normal incidence, the frequency of diffraction onset -- that is, at the lowest order -- is~\cite{Grann,Bruckner:07}
\begin{equation}
    \nu_{d} = \frac{c}{p(n+\sin{\theta_i})},  
\label{eq:diffraction}
\end{equation}
providing qualitative agreement with the RCWA calculation. We do not expect exact quantitative agreement because the RCWA calculation assumes the full average shape information for the two samples, as given in Table~\ref{tab:geometrical_measurement_parameters_table}, whereas Equation~\ref{eq:diffraction} only assumes a periodic array of scattering centers.

{}

\section{Conclusions}
\label{sec:conclusion}

SWS have advantages as ARC because they obviate the need for multiple materials and glues with precisely tuned indices of refraction, and because the index gradient produced can be smoother and tailored to specific applications. SWS are superior for cryogenic applications, because there is no need to match materials with different coefficients of thermal expansion.  Fabricating SWS on alumina and sapphire -- materials that have favorable optical properties in the millimeter and sub-millimeter wave band -- using standard machining approaches has been challenging because both materials are among the hardest available. They rank 9/10 on the relative hardness Mohs scale, and sapphire (alumina) has a value of 2500 (2000) HV on the Vickers hardness scale. In this paper we have extended our development of laser ablation as a tool to fabricate SWS for the millimeter and sub-millimeter band. We demonstrated structures on sapphire with aspect ratio $a = 3.7$. The newly fabricated SWS give a working bandwidth of 130\% with transmission above 90\% centered on 97~GHz, and of 116\% with transmission of at least 97\% centered on 102~GHz, arguably the largest bandwidths yet demonstrated in this wavelength range.

Using shorter pulse duration higher power laser and a more efficient fabrication process we have accelerated the AVRR for sapphire by a factor of 18 to 1.6 mm$^{3}$/min. Further acceleration is achievable with increase in laser power~\cite{Neuenschwander,boerner2019} and improvements in fabrication efficiency.

Although the native material was  non-birefringent we found shape asymmetries in the fabricated SWS. The origin of the asymmetries is yet to be understood. However, calculations show that proper relative alignment of samples reduces the magnitude of induced instrumental polarization due to differential reflection by a factor of 35 to 0.07\% at normal incidence, and to less than 0.6\% for incidence angles up to 20 degrees. Measurements at normal incidence are consistent with these predictions as they show the expected reduction in instrumental polarization, but are currently limited to an upper limit of 1\%.

These results and newer ones showing even higher ablation rates indicate that laser ablation of SWS on sapphire and on other hard materials such as alumina is an effective way to fabricate broad-band ARC; the technique has particularly strong advantages in the case of cryogenic applications.

\begin{acknowledgments}
We acknowledge the World Premier International Research Center Initiative (WPI), MEXT, Japan for support through Kavli IPMU. This work was supported by JSPS KAKENHI Grant Numbers JP17H01125, 19K14732, 18J20148, 18KK0083, and JSPS Core-to-Core Program, A. Advanced Research Networks. This work was also supported by the New Energy and Industrial Technology Development Organization (NEDO) project “Development of advanced laser processing with intelligence based on high-brightness and high efficiency laser technologies” by Council for Science, Technology and Innovation (CSTI), Cross-ministerial Strategic Innovation Promotion Program (SIP), “Photonics and Quantum Technology for Society 5.0” and by the Center of Innovation Program, from Japan Science and Technology Agency, JST. 
\end{acknowledgments}

\section*{DATA AVAILABILITY}
The data that support the findings of this study are available from the corresponding author upon reasonable request.

\bibliography{aipsamp}
\bibliographystyle{unsrt}

\end{document}